\documentstyle[pra,aps]{revtex}
\begin{document}
\draft
\author{V.B. Svetovoy\thanks{%
E-mail: svetovoy@nordnet.ru} \ and M.V. Lokhanin}
\address{Department of Physics, Yaroslavl State University, \\ Sovetskaya 14,
Yaroslavl 150000, Russia}
\title{Linear in temperature correction to the Casimir force}
\date{}
\maketitle

\begin{abstract}
We discuss the temperature correction to the Casimir force between nonideal
metallic bodies which caused disagreement in the literature. A general
method to find the troubling term is proposed that does not require a direct
reference to the Lifshitz formula. The linear in temperature correction is
shown to survive for nonideal metals. It is important for small separations
between bodies tested in the recent experiments.
\end{abstract}

\pacs{ 12.20.Ds, 03.70.+k }

%\newpage

The Casimir force \cite{Casimir} has been measured with high precision in
recent experiments \cite{Lam1,MR,RLM,HCM} between metallized sphere and
plate. Also there are plans \cite{Long,Fisch} to look for very weak
hypothetical forces where the Casimir force is the main background. All this
makes the precise evaluation of the Casimir force an important problem. Here
we will discuss a particular problem concerning the temperature dependence
of the force between macroscopic bodies made of real metals. Discussion of
this problem in the literature revealed disagreement between different
authors and at the moment there are three different results.

For perfect conductors the temperature correction has been found many years
ago \cite{Meh,Brown,Schw} and it is small for small separations between
bodies $a\ll c\hbar /kT$ or equivalently for low temperature. For a sphere
above a plate the leading term behaves as $\left( T/T_{eff}\right) ^3$,
where $kT_{eff}=\hbar c/2a$. This result follows from a general expression
for the Casimir force given by Lifshitz \cite{Lif,LP} modified for the case
of sphere-plate geometry with the proximity force theorem (PFT) \cite{PFT}:

\begin{equation}
\label{base}F(a)=-\frac{kTR}{4a^2}{\sum\limits_{n=0}^\infty {}}^{\prime
}{}\int\limits_{x_n}^\infty dxx\ln \left[ \left( 1-G_1e^{-x}\right) \left(
1-G_2e^{-x}\right) \right] , 
\end{equation}

\noindent where $R$ is the sphere radius,

$$
G_1=\left( \frac{x-s}{x+s}\right) ^2,\quad G_2=\left( \frac{\varepsilon
\left( i\zeta _n\right) x-s}{\varepsilon \left( i\zeta _n\right) x+s}\right)
^2,\quad 
$$

\begin{equation}
\label{xn}s=\sqrt{x_n^2\left( \varepsilon \left( i\zeta _n\right) -1\right)
+x^2},\quad x_n=\frac{2\zeta _na}c,\quad \zeta _n=\frac{2\pi nkT}\hbar 
\end{equation}

\noindent $\varepsilon \left( i\zeta _n\right) $ is the dielectric function
of the used material at imaginary frequencies. The prime over the sum sign
indicates that the first term $n=0$ has to be taken with the coefficient $%
1/2 $. Throughout this paper we will work with the force $F(a)$ between
sphere and plate. The force for the plate-plate configuration can be
restored as $F^{\prime }$ $\left( a\right) /2\pi R$.

For small temperature the sum in (\ref{base}) can be replaced by the
integral and the resulting force does not depend on the temperature at all.
This limit has been considered \cite{Lam2,LR} to calculate the force in the
conditions of experiments. In general, the replacement is true with the
precision $\sim T/T_{eff}$. For the atomic force microscope (AFM)
experiments \cite{MR,RLM} the smallest separation was $0.1\ \mu m$ and the
replacement error can be as large as 3\% at $T=300^{\circ }\ K$. It exceeds
the experimental errors $\sim $1\% and, therefore, the finite temperature
effect has to be taken into account. We define the temperature correction $%
\Delta _TF$ as the difference between forces written as the sum over $n$ and
as the integral instead of this sum.

The physical discussion of the first term in the sum (\ref{base}) raised
heated debate in the literature. It has a long history. For ideal metals the
temperature correction found from the Lifshitz formula \cite{DLP} did not
agree with that found with different methods \cite{Meh,Brown}. Schwinger 
{\it et al.} \cite{Schw} clearly demonstrated that when the static limit was
correctly taken in the Lifshitz theory the discrepancy disappeared. The
formulated condition is known as the Schwinger prescription: $\varepsilon $
should approach infinity before the frequency is allowed to go to zero.

Different results have been found for the $n=0$ term in the recent works
where the Casimir force was calculated between nonideal metals. All the
results one can parametrize \cite{LamT} with the following relation

\begin{equation}
\label{zterm}F_{n=0}\left( a\right) =\alpha \frac{kTR}{4a^2}\zeta \left(
3\right) , 
\end{equation}

\noindent where $\zeta \left( m\right) $ is the zeta-function and $\alpha $
is the parameter. In our paper \cite{SL1} $\alpha =1$ was found using the
Drude model for $\varepsilon \left( i\zeta \right) $ and the Schwinger
prescription. Doubt in applicability of the prescription for real metals was
expressed in \cite{BS} where straightforward calculation gave $\alpha =1/2$.
Using the plasma model for the metal dielectric function, $\alpha
=1-2c/a\omega _p$ was found \cite{BGKM} (see also \cite{GLR}), where $\omega
_p$ is the plasma frequency. Physical arguments based on the finite size of
plates in real experiments have led Lamoreaux to the result $\alpha =1$ \cite
{LamT}. There is no disagreement for the other terms in the sum (\ref{base}%
), so different $\alpha $ are responsible for different temperature
corrections. In this letter we will try to clarify the problem.

The $n=0$ term (\ref{zterm}) gives the classical contribution to the force ( 
$\hbar $-independent). At large separations between bodies ($T_{eff}\ll T$)
this is the only term in the sum (\ref{base}) which survives. All the other
terms are exponentially suppressed. In this sense one can say that (\ref
{zterm}) describes the contribution of the long wavelength or low
frequencies fluctuations (the static limit). Below we demonstrate that the
value of $\alpha $ can be found for real metals from very general analysis
without direct reference to the Lifshitz formula (\ref{base}) if one knows
its value for ideal metals. It is important because the static limit
calculated from (\ref{base}) is controversial: it gives different values for 
$\alpha $ if the metal is described by the plasma or the Drude models and
there is no way to reconcile the values in the limit of zero relaxation
frequency.

The force in the classical limit (\ref{zterm}) with an arbitrary parameter $%
\alpha $ can be written using only the dimensional analysis. More exactly
one can reproduce the force between two plates and then apply PFT. The value
of $\alpha $ can depend on the material parameters which define the metal
dielectric function $\varepsilon \left( i\zeta \right) $ at low frequencies.
In this limit $\varepsilon \left( i\zeta \right) $ depends on the only one
parameter which is the metal conductivity $\sigma $ or equivalently the
resistivity $\rho =1/\sigma $. The dimensionless parameter $\alpha $ can be
a function of the only one dimensionless variable

\begin{equation}
\label{alpha}\alpha =\alpha \left( \frac{c\rho \varepsilon _0}a\right) , 
\end{equation}

\noindent where $\varepsilon _0$ is the free space permittivity. For ideal
metals $\rho \rightarrow 0$ and $\alpha \left( 0\right) =1$ as was
established many years ago \cite{Meh,Brown,Schw}. Let us estimate the length
scale $L=c\rho \varepsilon _0$ for real good metals. Using the typical
resistivity $\rho \leq 10\ \mu \Omega \cdot cm$ one finds $L\leq 3\ \AA $.
Of course, one can expand the function $\alpha \left( L/a\right) $ in the
powers of $L/a$ but there is no too much sense in this. The reason is that $L
$ is a microscopic scale. Really, it is closely connected with the screening
length $L_s\sim {\it v}_F/\omega _p$ of degenerated electron gas, where $%
{\it v}_F$ is the Fermi velocity of electrons. If $\delta =c/\omega _p$ is
the field penetration depth and $l$ is the mean free path for electrons,
then $L\sim L_s\left( \delta /l\right) $. For good metals $\delta \sim l$
and $L$ is the same order as $L_s$. Taking into account such a short scale
will be excessive for the macroscopic Casimir force. Therefore, we can
conclude that the force in the static limit is the same as for ideal metals,
i.e. $\alpha =1$. Intuitively it seems quite obvious. In the static limit
there is no difference between metals because the free charges manage to
follow the slowly varying field. That is why the boundary conditions in this
case do not include any characteristic of a particular metal. This
conclusion means that if we try to calculate the force using the Lifshitz
formula (\ref{base}), the Schwinger prescription has to be applied to the $%
n=0$ term independently if it is a perfect or real metal. Lamoreaux \cite
{LamT} arrived at the same conclusion from quite different physical
consideration.

In ref. \cite{BS} fair calculation of the $n=0$ term in (\ref{base}) gave $%
\alpha =1/2$. This value does not depend on any material parameter and,
therefore, should be the same as for the ideal metal. However, we know that
for a perfect conductor $\alpha =1$ due to the Schwinger prescription. So
the disagreement has the same nature as in the old story described above.

In ref. \cite{BGKM} the plasma model was used for the metal dielectric
function

\begin{equation}
\label{pm}\varepsilon \left( i\zeta \right) =1+\frac{\omega _p^2}{\zeta ^2}. 
\end{equation}

\noindent This model gives good description of a metal in the infrared range
but cannot be used at low frequencies where $\varepsilon \left( i\zeta
\right) =1+(\varepsilon _0\rho \zeta )^{-1}$. Therefore, it will be good
everywhere except of the $n=0$ term for which one has to use the Schwinger
prescription. It is no wonder then that in the static limit a significant
correction to $\alpha =1$ has been found \cite{BGKM}: $\alpha =1-2c/a\omega
_p$. That is because in the plasma model the length scale $L=c/\omega _p$
(penetration depth) becomes macroscopic $L>150\ \AA $ and expansion in $L/a$
has sense.

The plasma model cannot be used at low frequencies even in principle. The
only way to get $\varepsilon \left( i\zeta \right) $ is the dispersion
relation

\begin{equation}
\label{dr}\varepsilon \left( i\zeta \right) =1+\frac 2\pi \int\limits_0^%
\infty d\omega \frac{\omega \varepsilon ^{\prime \prime }\left( \omega
\right) }{\omega ^2+\zeta ^2}, 
\end{equation}

\noindent where $\varepsilon ^{\prime \prime }\left( \omega \right) $ is the
imaginary part of the dielectric function at real frequencies. In the plasma
model there is no dissipation and $\varepsilon ^{\prime \prime }\left(
\omega \right) =0$. Eq. (\ref{pm}) one can derive from (\ref{dr}) only
calculating the integral with some small but finite relaxation frequency $%
\omega _\tau $. For finite dissipation always there is such a frequency for
which $\varepsilon \left( i\zeta \right) =1+(\varepsilon _0\rho \zeta )^{-1}$
with an arbitrary small $\rho $.

The authors \cite{BGKM} claim that some identical transformation of the
Lifshitz formula removes the controversy in the force for the plasma and
Drude models without the use of Schwinger prescription. This statement is
definitely wrong because any identical transformation cannot change the
result. Below it will be demonstrated explicitly.

Let us discuss now the temperature correction to the Casimir force. The sum
for $n>0$ in (\ref{base}) as a function of temperature contains a piece
linear in $T$ which exactly cancels for ideal metals the $n=0$ term giving
in the small temperature limit $T/T_{eff}<1$ the well known result \cite
{Brown}

\begin{equation}
\label{FT}F_T(a)=F_0(a)\left[ 1+\frac{45\zeta \left( 3\right) }{\pi ^3}%
\left( \frac T{T_{eff}}\right) ^3-\left( \frac T{T_{eff}}\right) ^4\right] , 
\end{equation}

\noindent where $F_0(a)=\pi ^3\hbar cR/(360a^3)$ is the bare Casimir force
between sphere and plate.

If we are using the dielectric function of a real metal, the cancellation of
the first term in (\ref{base}) is incomplete and the linear in $T$
contribution survives. That was noted first in \cite{SL1}, where eq. (\ref
{base}) with $\alpha =1$ was used for numerical evaluation of the Casimir
force. It was found that for the AFM experiments \cite{MR,RLM} the
temperature correction at the smallest separation is $4\ pN$ against the
experimental errors $1\ pN$. We can give now an explicit expression for the
correction.

In ref. \cite{BGKM} the authors did not use the Schwinger prescription in
the $n=0$ term and took the plasma model for $\varepsilon \left( i\zeta
\right) $. For this case they convincingly demonstrated that there was no
the linear in $T$ correction and the leading correction is only $\left(
T/T_{eff}\right) ^3$. As was shown above the prescription must be applied to
the $n=0$ term so the only difference in evaluation of the sum comes from
this term. This difference is just the correction we are looking for

\begin{equation}
\label{delT}\Delta _TF=\frac{kTR}{4a^2}\left\{ \zeta \left( 3\right) +\frac 1%
2{}\int\limits_0^\infty dxx\ln \left[ \left( 1-G_1e^{-x}\right) \left(
1-G_2e^{-x}\right) \right] \right\} +O\left( \left( T/T_{eff}\right)
^3\right) . 
\end{equation}

\noindent The integral here can be interpreted as the linear in $T$ term
contained in the sum (\ref{base}) for $n>0$ and, of course, it can depend on
the material parameters because the summation is going over nonzero
frequencies $\zeta _n$. On the other hand, since this integral appeared as
the $n=0$ term in (\ref{base}), we should take the functions $G_{1,2}$ at $%
x_n=0$. In this limit $G_2=1$ but $G_1\neq 1$. Using then the relation $%
\int_0^\infty dxx\ln \left( 1-e^{-x}\right) =-\zeta \left( 3\right) $ one
finds the final expression for the correction linear in $T$:

\begin{equation}
\label{delTf}\Delta _TF=\frac{kTR}{8a^2}\left[ \zeta \left( 3\right)
+\int\limits_0^\infty dxx\ln \left( 1-G_1e^{-x}\right) \right] +O\left(
\left( T/T_{eff}\right) ^3\right) , 
\end{equation}

\noindent where

$$
G_1=\left( \frac{x-\sqrt{x^2+\beta ^{-2}}}{x+\sqrt{x^2+\beta ^{-2}}}\right)
^2,\qquad \beta =\frac c{2a\omega _p}. 
$$

\noindent Let us stress that (\ref{delTf}) is true only for the plasma
model. When $\omega _p\rightarrow \infty $ the correction disappears as it
should be. Expansion in powers of $\beta $ gives

\begin{equation}
\label{exp}\Delta _TF=\frac{kTR}{8a^2}\zeta \left( 3\right) \cdot 8\beta
\left( 1-3\beta +O\left( \beta ^2\right) \right) +O\left( \left(
T/T_{eff}\right) ^3\right) . 
\end{equation}

\noindent For $\omega _p=2\cdot 10^{16}\ s^{-1}$ , $a=0.1\ \mu m$ and $%
R=100\ \mu m$ as in the AFM experiments \cite{MR,RLM} one gets $\Delta
_TF\approx 2.5\ pN$ using (\ref{delTf}) or calculating directly with the
help of (\ref{base}) and $2.9\ pN$ using (\ref{exp}). In the condition of
the Lamoreaux experiment \cite{Lam1} ($\omega _p=1.4\cdot 10^{16}\ s^{-1}$ , 
$a=0.6\ \mu m$ and $R=12.5\ cm$) the linear in $T$ part of the correction $%
\Delta _TF\approx 29\ pN$ is within the experimental errors. This conclusion
is in contrast with that of Bostr\"om and Sernelius \cite{BS} due to obvious
reason. If $\alpha =1/2$ as in ref. \cite{BS}, the $n=0$ term is not
canceled by the sum of those at $n>0$ and there is no an additional
suppression $\sim \beta $ as in (\ref{exp}).

The correction increases further if we will use the Drude dielectric function

\begin{equation}
\label{Drude}\varepsilon \left( i\zeta \right) =1+\frac{\omega _p^2}{\zeta
\left( \zeta +\omega _\tau \right) }. 
\end{equation}
\noindent In this case it has to be evaluated numerically using (\ref{base})
and subtracting the same formula with the integral instead of the sum. The
relaxation frequency $\omega _\tau $ influences mostly on the integral since
it changes low frequency behavior of the integrand. For typical value $%
\omega _\tau =5\cdot 10^{13}\ s^{-1}$ we found $\Delta _TF\approx 4.0\ pN$
in condition of the AFM experiments \cite{MR,RLM}. This value coincide with
that reported before \cite{SL1}. In the experiment \cite{HCM} the smallest
separation was $a=63\ nm$ and the absolute value of the correction is larger 
$\Delta _TF\approx 10\ pN$ \cite{SL2} though its relative value decreases.
It has to be compared with the experimental error $3.5\ pN$. Note that this
correction make the agreement between theory and experiment better \cite{SL2}%
.

In ref. \cite{BGKM} the authors claim that the Drude model has a principal
drawback which is discontinuity of $G_1$ at $x=0$ for the $n=0$ term.
Namely, $G_1=1$ at $x=0$ but $G_1=0$ at $x\neq 0$ in the limit $%
x_0\rightarrow 0$. In this connection we would like to stress that the
finite discontinuity in one point has no effect on the integral in (\ref
{base}). As was explained above the problem with the $n=0$ term has the
physical nature. One has to accurately takes into account the static limit
for which $\varepsilon \rightarrow \infty $ must be taken before allowing $%
\zeta \rightarrow 0$ for both real and ideal metals.

The problem with the $n=0$ term was declared \cite{BGKM} to be connected
with the discontinuity of $G_1$. This has led the authors \cite{BGKM} to the
wrong conclusion that the problem can be solved by the mathematical methods.
They transformed eq. (\ref{base}) with the Poisson formula and represented
it as the Fourier transformation

\begin{equation}
\label{four}F(a)=-\frac{kTR}{2a^2}{\sum\limits_{m=0}^\infty {}}^{\prime
}{}\int\limits_0^\infty dt\cos \left( 2\pi mt\right) \varphi \left( \tau
t\right) , 
\end{equation}

\noindent where

$$
\varphi (z)=\int\limits_z^\infty dxx\ln \left[ \left( 1-G_1e^{-x}\right)
\left( 1-G_2e^{-x}\right) \right] ,\quad \tau =2\pi \frac T{T_{eff}}. 
$$

\noindent It was stated that in this form the Lifshitz formula had no
problem for the Drude model. However, the Poisson formula is just an
identical transformation for the series and it cannot give any new result.
As we know without the Schwinger prescription it follows from (\ref{base})
that $\alpha =1/2$ \cite{BS}. The same value one can get from (\ref{four}).
It was demonstrated numerically \cite{Bost} and easy to do analytically.

To see it explicitly, let us consider the static limit $\tau \gg 1$.
Calculating the sum in (\ref{four}) for finite but large $M\gg \tau $ one
finds

\begin{equation}
\label{int1}F(a)=-\frac{kTR}{4a^2\tau }\int\limits_0^\infty dz\frac{\sin \pi
\left( 2M+1\right) z/\tau }{\sin \pi z/\tau }\varphi \left( z\right) . 
\end{equation}

\noindent The integral is saturated at $\pi z/\tau \sim 1/(2M+1)\ll 1$ and
we can replace $\sin \pi z/\tau \approx \pi z/\tau $. Then (\ref{int1}) can
be represented as

\begin{equation}
\label{int2}F(a)=-\frac{kTR}{4\pi a^2}\int\limits_0^\infty dy\frac{\sin y}y%
\varphi \left( \frac{\tau y}{\pi \left( 2M+1\right) }\right) . 
\end{equation}

\noindent We get the expected result from (\ref{int2}) taking the limit $%
M\rightarrow \infty $:

\begin{equation}
\label{int3}F(a)=-\frac{kTR}{8a^2}\int\limits_0^\infty dxx\ln \left[ \left(
1-G_1e^{-x}\right) \left( 1-G_2e^{-x}\right) \right] \quad for\ \tau \gg 1. 
\end{equation}

\noindent It coincide with the $n=0$ term in (\ref{base}) as it should be.
Therefore, using the Drude model we will get $\alpha =1/2$ from (\ref{four})
so as from the original Lifshitz formula (\ref{base}).

In conclusion, we have considered the linear in temperature correction to
the Casimir force at low temperatures or equivalently at small separations.
Special care has to be taken to get the contribution of the fluctuations in
the static limit ($n=0\ $term). This contribution is canceled for ideal
mirrors but cancellation is incomplete for real metals. A general method
based on the dimensional analysis has been proposed to calculate the
troubling $n=0$ term without direct reference to the Lifshithz formula. On
this bases it was shown that the Schwinger prescription had to be applied in
the static limit independently if it is real or ideal metal. The static
contribution is the same for the plasma and Drude models. The temperature
correction is shown to be important for the AFM experiments \cite{MR,RLM,HCM}
at the smallest separations but is less than the experimental errors for the
torsion pendulum experiment \cite{Lam1}.

\end{document}